# ProofPeer - A Cloud-based Interactive Theorem Proving System
by Steven Obua

## Introduction

Interactive theorem proving (ITP) has a long history. Most of that history, starting roughly around 1970, can be traced back to three systems: Automath [1], Edinburgh LCF [2], and Mizar [3].

Probably the most popular ITP systems these days are Isabelle [4a] and Coq [5]. Isabelle is a descendant of Edinburgh LCF. Its declarative extension Isabelle/Isar [4b] has been heavily influenced by Mizar. Coq can be viewed as a descendant of the Automath system in that both systems are based on the Curry-Howard correspondence [6] (a proof is a program, the formula it proves is a type for the program). Coq is implemented in OCaml [7], which is a dialect of Standard ML [8] which is one of the results of the Edinburgh LCF project, therefore Coq has been heavily influenced by Edinburgh LCF as well. Recently Coq has also gained declarative features similar to those found in Mizar and Isabelle/Isar.
Another ITP system is HOL-light, another Edinburgh LCF system which has a small and actually formally verified kernel [17].

So ITP has been around for over 40 years now and there have been many notable successes and ongoing efforts, among them:

- Formal specification of the ARM instruction set and verification of the ARM6 micro-architecture by Anthony Fox in 2002 [10]
- Formal proof of the prime number theorem by Jeremy Avigad in 2004 [16]
- Formal proof of the Jordan curve theorem by Thomas Hales in 2005 [18]
- Formal proof of the Four Color Theorem by George Gonthier et al. in 2005 [13]
- Formal verification of a compiler back-end by Xavier Leroy in 2006 [11]
- Formal verification of the seL4 operating system microkernel by Gerwin Klein et al. in 2009 [12]
- Formal proof of the Kepler Conjecture by Thomas Hales et al., ongoing effort since 2003 [14]

- Formal classification of simple finite groups by George Gonthier et al., ongoing effort since 2006 [15].

Furthermore, automated theorem proving techniques promise [19] to reduce a lot of the drudge work that is associated with formal theorem proving.

Therefore one can conclude that in principle, ITP technology is ready for prime time. So why is it that the vast majority of mathematicians and engineers have not used an ITP system for their work, ever? Most often, they are not even aware that ITP technology exists and is out there, ready to be used.

Gonthier, among others, proposes that this is due to the *library problem*. To prove a non-trivial mathematical proof, usually it is necessary to have access to a large corpus of mathematical notions. To deal with these notions, one also needs appropriate syntax for them, previously proven theorems about them, and decision procedures and executable algorithms involving them. In short, one needs a large formalized math library. Once such a library has been built that is large enough, they will come.

While the library problem is definitely one of the main reasons why ITP has not seen wide-spread adoption so far, we think there are other major reasons and problems that need to be tackled as well. These issues are actually not separate from the library problem, but might very well interact with it in hard to quantify ways. Let us examine these issues:

*1. A mathematical theorem lives forever, but a **formalized** mathematical theorem is pretty fragile.*
A lot of the attraction of mathematics comes from the aura of eternity that surrounds it. Once you have proven a mathematical theorem, it will live on forever and ever. Even if you have *not* proven it yet, it can live forever at least as a conjecture.
The situation is very different with current ITP systems.
First, your formalized theorem might crumble under the change from one version of your ITP system to the next one. During the Verisoft project, a large-scale project that aimed to verify a complete computing stack down from digital circuits all the way up to applications such as an email client, people could not switch from the version of their ITP system that they used in the beginning, Isabelle 2005, to a newer version like Isabelle 2010. Such a change would have had many benefits like improved automation [19], but could not be done because a lot of minor details of the Isabelle system had changed.

The second cause your formalized theorems crumble is that they depend on definitions that have been changed, or that their proof depends on other theorems that have changed. This sort of dependency usually becomes worse the more you need to interface with the formalizations of other people, but can already be very annoying if all the formalizations involved are your own.

*2. To do math, one just needs paper and pencil and one is ready to go. ITP systems instead usually require quite elaborate installation procedures, sometimes involving setting up a whole new operating system on your computer.*

A mathematician usually requires very little tools besides her brain and paper and pencil, and derives part of her fun and pride from this very fact.

On the other hand, all the major ITP systems require a quite specific setup involving some flavor of Unix, and some way of running ProofGeneral [42]. To most computer scientists, establishing this setup is rather trivial, but people interested in theorem proving coming from another background, like many mathematicians, struggle with such a setup. And even people that are in principle equipped to handle such a setup might refrain from it just because of inertia.

*3. How can the engineer make use of the formal artifacts she created? How easy is it to create new tools and apply tools that interact with these formal artifacts?*

Formalizing and verifying a theorem or an algorithm usually involves some extra work (which often also comes with easily neglected extra benefits like hightened insight), so how can an engineer justify this extra work? For an engineer, usually a formalization is not the end point of her work, but a stepping stone along the way to the final product. Sometimes it is possible to make reuse of this formalization stepping stone in other phases of the product creation. To take advantage of such a possibility, custom-built third-party tools need to be able to access various parts of the formalization. Existing ITP systems sometimes provide ways for import / export to other ITP systems, but usually they provide no standard way to easily access and maybe even modify a formalization from outside the ITP system.

*4. How easy is it to share formalized artifacts between different users of the system? How can a user discover existing formalizations relevant to her? How can elegant and difficult formalizations be distinguished from trivial and messy ones? How are people rewarded that contribute to the system?*

Sharing your formalization with other people is something that can apparently be done easily with existing ITP systems. Just send them your proof document via email. Some systems go even beyond this: for example, Isabelle provides the *Archive of Formal Proofs* (AFP) [21]. You can submit your proof document to the AFP, and after a short review phase it becomes part of the AFP. There is a central location where you can look at the accepted proof documents that form the AFP and from where you can retrieve them if you want to make them part of your own formalization. Once a proof document has been accepted by the AFP, it is taken care of that any problems due to a version change of the Isabelle system are fixed. While obviously the AFP is already a big step in the right direction, and obviously superior in many respects to sharing a proof document casually via email, in its current form it does not scale. The bottleneck is the review phase, and the need to manually fix broken proof documents when going from one version of the ITP system to another. As long as the AFP has the size of a typical journal, it remains manageable. But imagine an environment in which there are 1000 or more submissions each day. Then clearly the AFP model breaks down.
Once such an archive of proof documents reaches a certain size, there is the question of how to find those proof documents I might be interested in? Can the question of whether a proof document is worth to become a part of the archive be decentralized? And finally, if somebody is making high-value contributions to the archive on a regular basis, how can she be rewarded and encouraged?

Our proposal is to attack all of these issues in a unified fashion by building a cloud-based ITP system which we call *ProofPeer*. The main advantage of being cloud-based is that the ITP system is centralized, but scalable. In the following we describe the architecture of ProofPeer, the design decisions we have made so far, and those design decisions we will have to make in the future.

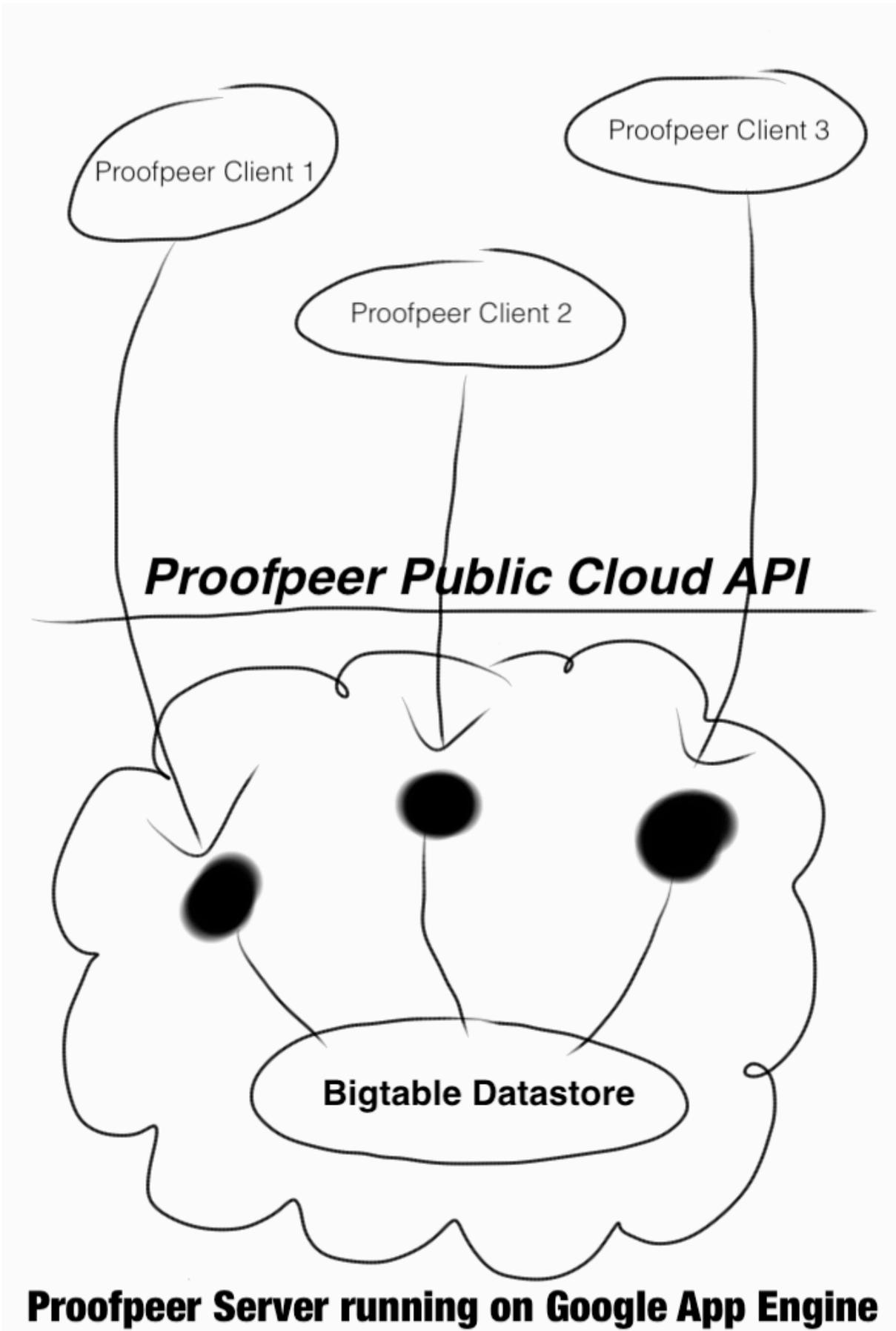

# High-level System Architecture of ProofPeer

The high-level system architecture of ProofPeer is similar to that of many modern cloud-based applications. The ProofPeer server is running on Google App Engine (GAE) [22] and can be accessed at http://proofpeer.appspot.com. GAE is a platform that provides scalable computing, caching and persistence resources. Using GAE has the advantage that no system administration like manual setup and configuration of server machines is necessary. Instead, GAE provides various models of how scalable applications can be run on it. We have chosen to run ProofPeer via the Java [23] platform model. This means that the ProofPeer server is essentially a collection of Java servlets [24] that communicate with each other only through distributed persistent storage in the form of Google's BigTable database [25] using the objectify abstraction [26], and through a distributed caching service called MemCache [27]. The advantage of using the Java platform model is that the server can be programmed in Scala [28], which is a language that has its origins in academia but is also used at places like Twitter [29]. Scala combines functional programming features with advanced object oriented features and expressive static typing, and is thus well-suited for the implementation of ITP systems.

Users of the ProofPeer system interact with the ProofPeer server via a ProofPeer client. A ProofPeer client interacts with the ProofPeer server via the ProofPeer public cloud API.

Every functionality of the ProofPeer server is exposed via this API. ProofPeer comes with a built-in client that is programmed using standard web techniques like Javascript, CSS and HTML. This client can be accessed by pointing a modern web browser (more specifically: Chrome, Safari 4+, IE 9+, Firefox 3.5+) to http://proofpeer.com (which redirects this request to http://proofpeer.appspot.com). This means that ProofPeer can be used without any installation hassles by anyone who has access to a reasonably recent web browser.

## Public Cloud API

Because the built-in ProofPeer client interacts with the ProofPeer server only via the public cloud API, anyone can build an alternative client or a tool that accesses the ProofPeer server. The API is designed in such a way that the consistency of the

ProofPeer server is always guaranteed. Specifically this means that no matter how you access the server via the API, you can only create valid theorems, never an invalid one.

The API allows the retrieval and manipulation of various entities. Currently there are the following main kinds of entities:

- *users* (also called *proofpeers*)
- *sessions*
- *contexts*
- *types, terms and theorems*
- *proofscripts*
- *chronicles*

All of these entities are not only conceptually important, but also have physical manifestations that can be stored to and retrieved from persistent storage. In the following we will look at all of these kinds of entities.

**Users and Sessions**

Every *user* of ProofPeer is uniquely identified by her login. To interact with most of the functions of the API, a user first needs to establish a *session*. To do so, the user sends her login together with her password to the server and receives a session id. In the following interactions with the API, the user does not provide anymore her login and password, but the session id. The user can quit a session anytime by sending a logout command to the server. At this point the session id becomes invalidated and cannot be used anymore for further API interactions. At any point in time, there is at most one session associated with a user. If a user requests a session without having ended a previous one, she simply joins the previous one.

**Monomorphic Higher-Order Logic Set Theory**

The logic of ProofPeer is monomorphic higher-order logic set theory (HOL-ST). *Polymorphic* higher-order logic set theory has been used before in interactive theorem proving by Gordon [30] and also by Obua [31]. Polymorphic HOL-ST is simply-typed polymorphic higher-order logic, enriched by an additional type *set* that models, with the help of additional axioms, the universe of Zermelo-Fraenkel sets. There is actually no need to explicitly add an axiom of choice for ZF, as this axiom is a consequence of the properties of the Hilbert choice operator, which is already part of higher-order logic.

The reason to choose HOL-ST is that set theory provides a widely accepted and flexible foundation of mathematics. On the other hand, higher-order logic has been used most successfully in previous applications of higher-order logic. Merging higher-order logic with set theory will hopefully allow to unite many of the benefits of both approaches.

The reason why monomorphic HOL-ST has been chosen over polymorphic HOL-ST is that in earlier experiments with HOL-ST, it turned out that there is often a conflict how to formalize a certain notion. For example, should the natural numbers be formalized as a set, or as a type? To avoid these conflicts, in ProofPeer it is not possible to define new types (at least not in the sense of higher-order logic), new notions will therefore automatically be defined as sets whenever possible. Without the ability to define new types, general polymorphism loses most of its appeal and is therefore excluded from ProofPeer. This has the additional advantage that defining constants locally is no problem in ProofPeer; this would lead to inconsistency if polymorphism where included [32].

**Contexts**

All theorem proving in ProofPeer is done in a given *context*. There is a predefined context called *root*, which is the starting point for any development that starts from scratch. Contexts acknowledge that mathematical reasoning is linear, therefore any context except the root context has exactly one parent context. The root context has no parent.

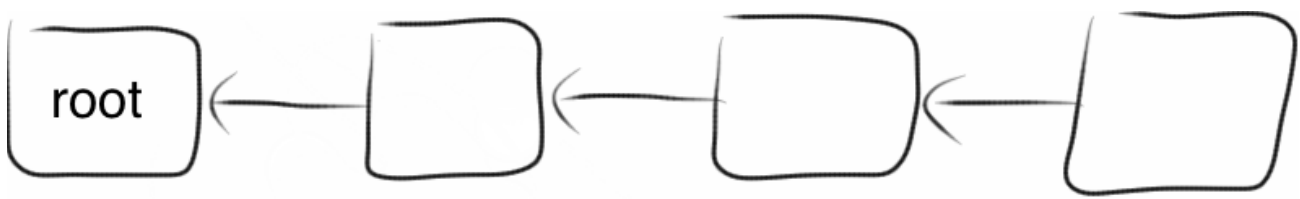

Apart from a reference to its parent context, a context also contains a reference to its *owner*. Furthermore, a context consists of four (possibly empty) components:
- a list C of logical constants introduced by this context;
- a list A of assumptions; these assumptions are axioms that may refer to the constant introduced in this context;
- a map V from names to values; the idea is that V binds certain names to certain values; these values can be terms, types, theorems, references to contexts; these values can also be booleans, integers, strings, vectors, lists, sets or maps; they can even be functions;
- a set U of names; the elements of U and the keys of V must be disjunct; the idea is that you can unbind a name by placing it in U;

These components work in a cumulative fashion; for example, in order to determine which logical constants are available in a certain context K, one has to consider the C component of K *and* the C components of the parent of K and so on.

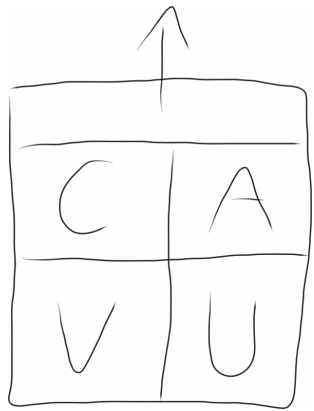

Before looking at the different kinds of contexts, we take a closer look at terms, types and theorems.

**Types**

The syntax for types $\tau$ is
$$\tau := \mathsf{set} \mid \mathsf{prop} \mid (\tau) \mid \tau_1 \to \tau_2$$
ProofPeer uses Unicode throughout. Nevertheless, it is possible to only use ASCII characters. The ASCII representation of "$\to$" is "->".

Note that, as explained earlier, there are no type variables, and there is no facility to define new types.

**Terms**

A term $\varsigma$ is either a built-in constant, a bound variable, a lambda expression, or an application. Only well-typed terms are allowed. The syntax for terms is:

$$\varsigma := c \mid x \mid \lambda x : \tau. \varsigma \mid \varsigma_1 \varsigma_2 \mid (\varsigma) \mid \varsigma : \tau$$

Note that variables are always bound, either by a lambda-expression, or by a logical constant in the context! Therefore terms, unlike types, make only sense relative to a given context, in particular, well-typedness is only defined relative to a given context.

Instead of

$$\lambda x : \text{set}. \varsigma$$

it is possible to just write

$$\lambda x. \varsigma$$

There are other forms of syntactic sugar like infix or quantifier notation that depend on the particular built-in constant. The built-in constants of ProofPeer are presented below. Note that although some constants are of polymorphic nature, only concrete monomorphic instances of them may appear in any term $\varsigma$.

| Constant | ASCII Representation | Type |
|---|---|---|
| = | = | $\tau \to \tau \to \text{prop}$ |
| $\forall$ | _all | $(\tau \to \text{prop}) \to \text{prop}$ |
| $\exists$ | _exists | $(\tau \to \text{prop}) \to \text{prop}$ |
| $\varepsilon$ | _choose | $(\tau \to \text{prop}) \to \tau$ |
| $\longrightarrow$ | --> | $\text{prop} \to \text{prop} \to \text{prop}$ |
| $\wedge$ | _and | $\text{prop} \to \text{prop} \to \text{prop}$ |
| $\vee$ | _or | $\text{prop} \to \text{prop} \to \text{prop}$ |

| Constant | ASCII Representation | Type |
|---|---|---|
| ¬ | _not | prop → prop |
| true | | prop |
| false | | prop |
| ∈ | _elem | set → set → prop |
| ∅ | _emptyset | set |
| 𝒫 | _powerset | set → set |
| ⋃ | _Union | set → set |
| ⋂ | _Intersect | set → set |
| ∪ | _union | set → set → set |
| ∩ | _intersect | set → set → set |
| ⊆ | _subset | set → set → prop |
| | _Singleton | set → set |
| | _Separation | set → (set → prop) → prop |
| | _Replacement | set → (set → set) → set |

**Equality of Terms**

Often in order to apply certain proof rules, ProofPeer needs to test if two terms are equal. General equality is undecidable, of course, so ProofPeer tests for structural equality after normalizing the terms. Currently normalization is done with respect to the usual conversions [33] of simply-typed lambda calculus plus a few additional simple normalizations:

- α-conversion
- β-reduction
- η-reduction

- negation (i.e. ¬ ¬ x reduces to x, ¬ true to false, ¬ false to true, a ⟶ false to ¬ a)

Actually α-conversion is not necessary because ProofPeer employs De Bruijn indices [34] to represent variables.

In future versions of ProofPeer this notion of equality is very likely to be extended, for example with simple computational equality.

**Syntactic Sugar for Terms**

The following table lists representative examples for what syntactical sugar is available for terms:

| **Syntactic Sugar** | **Translation** |
|---|---|
| ∀ x : τ. P | (∀ : (τ → prop) → prop) (λ x : τ. P) |
| x ∧ y | (∧) x y |
| {x} | _Singleton x |
| {x, y, z} | {x} ∪ ({y} ∪ {z}) |
| { x ∈ X \| P x } | _Separation X (λ x. P x) |
| {f x \| x ∈ X} | _Replacement X (λ x. f x) |

Currently, all syntactic sugar that can be used in ProofPeer is built-in and cannot be extended. It is an important research agenda to find safe ways for extending mathematical notation. Interesting aspects of what *safe* means in this context have been examined previously by Wiedijk under the notion of Pollack-inconsistency [35].

**Theorems**

A theorem consists of a proposition, i.e. a term that has type prop. In the tradition of Edinburgh LCF [2], theorems can only be constructed in such a way that their proposition is guaranteed to hold. There is always a context associated with a theorem. In particular, "false" can be a theorem relative to a context with contradicting assumptions.

**Kinds of Contexts**

The meaning of a context is determined by its parent and its components C, A, V and U. Given a parent context, there are currently 7 different ways to create a new context:

1. *fix* introduces a logical constant of name x and type $\tau$; therefore C is the singleton set $\{x : \tau\}$ and the other components are empty.
2. *assume* introduces a new axiom h, where h is a proposition, therefore A is the singleton set $\{h\}$. Let h' be the theorem with proposition h. Then $V = \{\text{fact} \to h'\}$ is the map that maps the name "fact" to the theorem h'. Optionally the user can provide a name n for h', then we have $V = \{\text{fact} \to h', n \to h'\}$ instead (we will leave out this comment in the following descriptions). The components C and U are empty.
3. *define* introduces a new logical constant of name x via definition by a term d, which has some type $\tau$. Then $C = \{x : \tau\}$, $V = \{\text{fact} \to (x = d)\}$ and A and U are empty.
4. *obtain* introduces logical constants x, y, ... , given a theorem with proposition $\exists$ x. $\exists$ y. ... p. Then $C = \{x, y, ...\}$, $V = \{\text{fact} \to p'\}$, where p' is the theorem with proposition p, and A and U are empty.
5. *have* takes a proposition p and a theorem p' and checks if the proposition of p' equals p. Then $V = \{\text{fact} \to p'\}$. All the other components are empty.
6. *bind* takes a name n and a value v. Then $V = \{n \to v\}$, and the other components are empty.
7. *unbind* takes a name n. Then $U = \{n\}$, and the other components are empty.

**The Context Tree**

Every context except the root context has exactly one parent. Therefore at any point in time, the set of all contexts in ProofPeer forms a tree.
Context A is an *ancestor* of context B if there is a non-empty chain of parents leading from B to A. In particular, root is an ancestor of every context except the root context. The *depth* of a context is the number of its ancestors.

## De Bruijn Notation

In order to easily shift terms and theorems from one context to another without the danger of name clashing, ProofPeer uses de Bruijn indices internally. Nevertheless, because the user uses names, there needs to be a way to deal with name clashes even though internally, everything is fine. Therefore ProofPeer allows to annotate variable names in terms with indices. An index of 0 refers to the last bind of the given name, an index of 1 refers to the next to last bind, and so on. It is also possible to leave out the name entirely and work with raw de Bruijn indices.

## Moving Terms Between Contexts

Much of the usefulness of contexts derives from the fact that there are relatively straightforward ways to define how to move terms and theorems from one context to another. Let us first look at how to move terms between a context A and a context B, where A is an ancestor of B.

ProofPeer uses de Bruijn indices to represent bound variables in terms. A variable in a term is either bound to the variable introduced by a lambda construct, or to a logical constant introduced by a context.

Any term t can be moved from context A to context B by adding k to those bound variables in t which refer to logical constants. Here k is the number of logical constants that have been introduced in B relative to A, i.e. the sum of the lengths of the C-components of all children of A on the path from A to B.

On the other hand, not every term t can be moved from context B to context A, because t might contain variables bound to logical constants which exist in context B, but not in context A. But it is easy to check using only k and t if t can be moved from B to A, and to actually move t if this turns out to be possible.

If neither A is an ancestor of B, nor B is an ancestor of A, let C be the context of maximum depth that is ancestor of both A and B. Then moving t from A to B means first moving t from A to C, resulting in t', and then to move t' from C to B.

## Moving Theorems Between Contexts

Let again context A be an ancestor of context B, and t a theorem in context A. Then t can be moved to B by just moving the proposition of t from A to B, because if t is true in A, then certainly it must also be true in B.

Unlike terms, theorems can *always* be moved from context B to context A. Of course in this case, it is not enough simply to move the proposition of t from B to A (which even might not be possible anyway). Instead we transform the theorem t by first moving t to the parent B' of B, resulting in a theorem t'. Then we move t' from B' to B" and so on, until we reach A.

Let us therefore now look at how to move a theorem from B to the parent of B. This depends on the components of B: C (logical constants), A (assumptions/axioms), V (binds names to values) and U (unbinds names). Let furthermore V' be the map that results from V by removing from it all bindings that bind a name to

- a theorem u such that the proposition of u is an element of A,
- or to a value that is not / does not contain a theorem (even in hidden form, like inside a function).

Let furthermore t be the proposition of a theorem in context B. Then the table below describes how the proposition of the moved theorem looks like.

| case | proposition of moved theorem |
|---|---|
| V' is empty<br>$C = [c_1, c_2, ..., c_n]$<br>$A = [a_1, a_2, ..., a_n]$ | $\forall c_1 c_2 ... c_n.\ a_1 \longrightarrow a_2 \longrightarrow ... \longrightarrow a_n \longrightarrow t$ |
| V' is not empty<br>$C = [c_1, c_2, ..., c_n]$<br>A is empty<br>$[d_1, d_2, ..., d_m]$ results from C by removing all $c_i$ that do not appear in t | $\exists d_1 d_2 ... d_m.\ t'$<br><br>where t' results from t by readjusting the bound variable indices accordingly |

All of the current seven kinds of contexts fall into one of the two categories outlined in above case-distinction table. In particular, no context has both non-empty V' and non-empty A.

**Storing Contexts**

Google App Engine comes with a high-performance, scalable, distributed datastore which is implemented on top of Google BigTable [25].

This datastore stores *entities*. An entity has a *kind*, which is used to categorize the set of all entities stored in the datastore. An entity consists of a key that uniquely identifies it. The entity furthermore has *fields*, which are used to store the information the entity contains. The number and names of its fields does not depend on the kind of the entity.

In order to store contexts into the datastore, an obvious choice is to introduce a new kind of entities such that each entity of this kind stores a context. To reduce the number of necessary datastore round trips, though, not only a single context can be stored per such entity, but an entire chain of contexts. Thus a stored context is uniquely identified by the key of its entity together with an index pointing out its position in the chain of contexts that is stored in the entity.

The table below lists the fields of a context entity.

| field name | field content |
|---|---|
| id | the key of this entity |
| parent_id | the key of the entity that contains the parent context of the first context stored in this entity |
| parent_index | the index of the parent context |
| owner | a reference uniquely identifying the owner of this entity |
| timestamp | encodes the time of when this entity has been stored |
| proofpeerversion | the version of ProofPeer that stored this entity |
| chain | an array of quintuples (k, C, A, V, U), where each quintuple encodes information about a single context; k is the kind of the context (fix, assume, ...), and C, A, V and U are the components of the context |

## Why Contexts?

The explicit introduction of contexts as first-class citizens that can be manipulated and stored just like terms and theorems, and the definition of explicit rules of how to move terms and theorems between them seems to be new in interactive theorem proving. Of course, contexts have been around for a long time in more or less implicit form in ITP systems that employ declarative forward-reasoning like Isabelle/Isar [4b]. The attractiveness of explicit contexts in a cloud-system like ProofPeer stems from the fact that contexts can function as a very fine-grained unit of theorem proving state that can be shared between users, internal ProofPeer processes, and external third-party tools. Once a context has been created, it is immutable. Sharing it can therefore be done without creating many of the conflicts that can otherwise be expected in concurrent settings.

Because contexts are relatively simple and basic, they can be expected to remain relatively unchanged during the life-time of the ProofPeer system, thus providing the stable backbone on which the intelligent versioning of ProofPeer rests.

## Chronicles

*Chronicles* enable the versioning of contexts. A chronicle has a name and is owned by a user / proofpeer. The login of the user and the name of the chronicle together uniquely identify a chronicle.

A chronicle manages a list of *chronicle versions*, sorted from newest to oldest. Each such chronicle version V is associated with a set S of contexts that are owned by the chronicle version. Out of these contexts S, one context $f \in S$ serves as the *final context* of V.

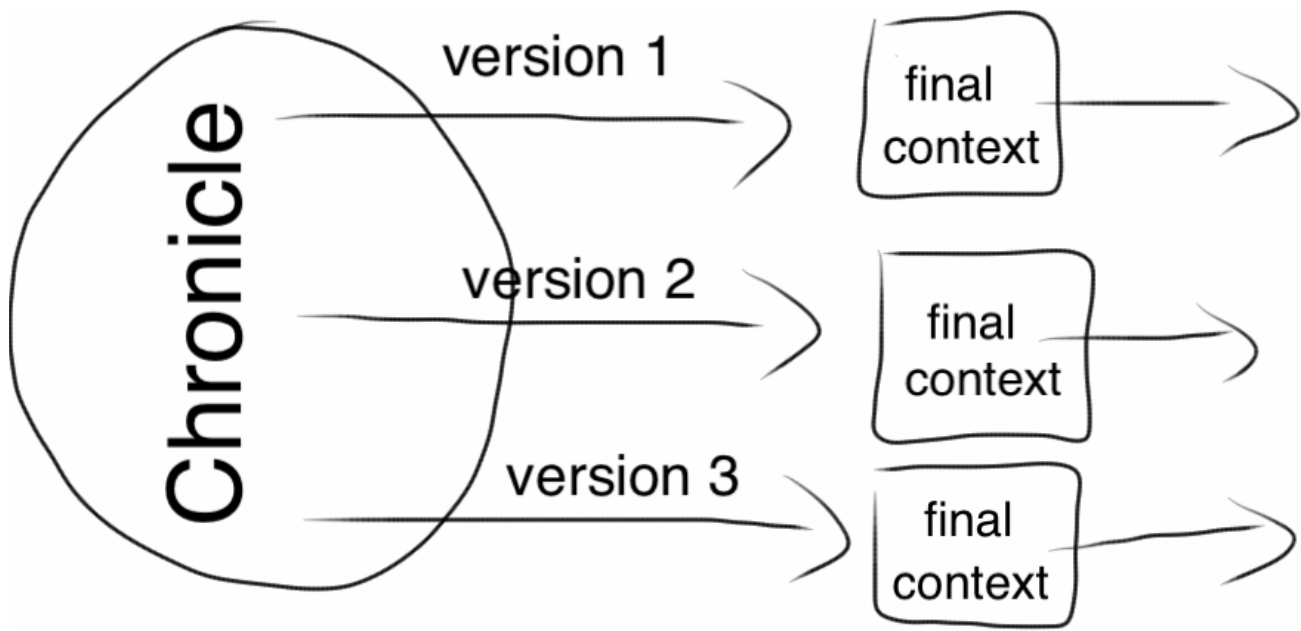

Given two chronicles A and B, and a chronicle version V of A, and a chronicle version W of B such that $V \neq W$, V is said to be *directly dependent* on W if one of the contexts owned by V has a parent context that is owned by W. If there is a chain of chronicle versions $U_1, ..., U_n$ (belonging to chronicles $C_1, ..., C_n$) such that V directly depends on $U_1$, $U_1$ directly depends on $U_2$, ..., and $U_n$ directly depends on W, then V is said to be *dependent* on W (in particular, direct dependency implies dependency). It is forbidden for a chronicle version to depend on a version of the same chronicle, so before creating a context ProofPeer needs to check that this constraint will not be violated.

Both notions of dependency can be extended from chronicle versions to chronicles: chronicle A is said to be (directly) dependent on B, if the newest version V of A is (directly) dependent on any version W of B. Because of the restriction we placed on dependency, it is never possible that a chronicle depends on itself.

Because the ProofPeer system itself can evolve over time, there is also a *root chronicle*. Actually, the root context is the final context of the newest root chronicle version.

A chronicle A is called *up-to-date* if its newest chronicle version only depends on other newest chronicle versions.

The notion of dependency induces directed acyclic connected graphs both on chronicles and chronicle versions, there being an edge from u to v if v is directly dependent on u.

The goal is to have only up-to-date nodes in the chronicle graph. To achieve this, the ProofPeer system tries to generate automatically the newest versions of chronicles that are not up-to-date. If this fails for a chronicle, the chronicle is flagged as *out-of-date*.

Note that out-of-date chronicles are still perfectly valid chronicles pointing to perfectly valid contexts which can be used even for new developments. This is what we mean by *intelligent versioning*: ProofPeer notices when chronicles fall out-of-sync, and tries to repair the damage automatically. This repair might succeed or not. In both cases, users don't have to worry that their existing and current developments are broken by out-of-date chronicles.

We have not mentioned so far exactly *how* ProofPeer tries to generate a new version of a chronicle. This is because in principle, one can think of many ways this can be achieved. All one has to do is to associate a chronicle with some method for generating contexts. The design of ProofPeer currently encompasses only one such method: proof scripts.

**ProofScript**

A *proof script* is a program for generating contexts. The syntax and semantics of ProofPeer proof scripts has been inspired by both Isabelle/Isar [4b] and Babel-17 [36]. The resulting language is called *ProofScript*.

The details of ProofScript will be described in another document, but we give an overview here.

ProofScript is dynamically typed. Nevertheless, each value in ProofScript has a specific type and adheres to the constraints and guarantees that this type implies. There is a type *theorem*, a type *context* and types *type* and *term*. There are also conventional types like *integer*, *string*, *list*, *vector*, *set*, *map* and *function*.

A ProofScript program consists of statements. A statement maps a pair (C, E), where C is a context and E is an environment, to another such pair (C', E').  Let us look for example at the *fix* statement:

**fix** s

Here s is an expression that should evaluate to one of the following things:

1. A *string* of the form "x : ty" , where x is the name of a new logical constant and ty denotes a type. Then C' has parent context C and fixes x to be a new logical constant of type ty. The environment does not change, therefore E' = E.
2. A *string* of the form "x ∈ D". where x is the name of a new logical constant, and D denotes a set. This actually generates an intermediate context C'' with parent C that fixes x to be a new logical constant of type set. Then C' is the context with parent C'' that assumes "x ∈ D". Furthermore E' = E ∪ {fact → h}, where h is a theorem in context C' with proposition "x ∈ D".
3. A pair (x, ty), where x has type *string* and ty has type *type*. This is just another form of case 1.
4. A pair (x, d), where x has type string, and d has type *term* and needs to be well-typed and having the logical type set. This is just another form of case 2.

Cases 2 and 4 above have an extended syntax in order to label the assumption:

$$\textbf{fix } n = s$$

This additionally binds n to be the name of the assumption, both in the context C' and in the environment E'.

All other kinds of creating a context also have their corresponding ProofScript statements.

Statements can be bundled as *blocks*. Typically a block starts with **begin** and ends with **end**, like in

```
begin
  fix "x ∈ N"
  fix "y ∈ N"
  assume "x ≠ y"
  ....
end
```

There are also statements that do not modify the context, but just the environment: **val** lets you introduce new non-recursive values into the environment, while **def** is used for mutually recursive definitions, e.g.

```
def even n = if n == 0 then true else odd (n - 1) end
def odd n = if n == 0 then false else even (n - 1) end
```

```
    val sevenIsOdd = odd 7
```

Besides **begin** ... **end** there are other control statements that can contain blocks:
- if ... then ... else ... end
- for ... in ... do ... end
- while ... do ... end
- match ... case ... => .... case ... => ... end

ProofScript has not only statements, but also *expressions*. The expression **root** denotes the root context, while the expression **this** always denotes the current context. An expression f can be applied to an expression g:

$$f\ g$$

Usually f will evaluate to a function for the above to make sense, but application makes not only sense for applications: an important case is when f evaluates to a theorem with proposition "h $\longrightarrow$ p" and g evaluates to a theorem with proposition "h", then f g evaluates to a theorem with proposition "p"; another case is when f evaluates to a theorem "$\forall$ x. p" and g evaluates to a term u that matches the type of "x", then f g evaluates to a theorem "q", where q results from p by substituting u for x.

Blocks are expressions, too. If the last statement of a block is actually not really a statement, but an expression, then the value of the block is the value of that last expression. If there are only proper statements in a block, then the value of the block is a reference to the context that the block created.

ProofScript has syntax for term literals. To denote a term, just enclose it in single-quotes, like in 'x $\in$ N'. Note that whenever a term is used, a string could be used as well; the difference is that the string must be converted first to a term before it can be used as a term.

It is possible to start the generation of contexts from an arbitrary context:

$$\textbf{with}\ c\ \textbf{do}\ b\ \textbf{end}$$

first evaluates c to yield a context C, and then executes the block b with C as a starting point.

**Have**

The *have* statement deserves special mention. It illustrates well how contexts are used to produce theorems. Its general syntax is

$$\textbf{have } n = \text{guard } \textbf{by } \text{thm}$$

where thm is an expression that evaluates to a value of type *theorem*, n is the name that the theorem is going to be bound to, and guard is a string / term that has to match the proposition of the theorem. If the guard does *not* match the proposition of the theorem, then the have statement fails and throws an exception.

Now, if thm does not evaluate to a theorem, but to a context C instead, ProofPeer tries to convert the context to a theorem by accessing C.fact. This makes sense because all the theorem producing statements automatically bind the produced theorem to the name "fact". The theorem C.fact is then moved into the same context as the *have*-statement, using the moving strategy explained earlier.

Let us look at an example:

```
have impl = "∀ x : prop. x ⟶ x" by
   begin
      fix "x : prop"
      assume h = "x"
      have "x" by h
   end
```

This proof script would execute successfully and result in binding impl to the theorem "∀ x. x ⟶ x".

**Identifier Resolution in ProofScript**

So far we have not talked much about how identifiers are resolved in ProofScript. An expression is evaluated relative to a pair (C, E), where C is a context, and E is the environment that keeps all current identifier bindings. So E contains all bindings that have accumulated via the V and U components of C and its ancestors, and on top of that those bindings that have been established during the execution of the proof script but are not stored permanently in a context.

Given an expression c that evaluates to a context C, you can access the accumulated bindings of C with the expression c.id, where id is the name of the particular binding you want to retrieve.

The proof script always runs relative to a certain assignment of chronicles to chronicle versions. Usually this assignment is just assigning to a chronicle its newest version. You can access the final context of the chronicle version that has been assigned to a chronicle with name n via @n. Note that chronicles are not uniquely identified by their name, but only by the combination of chronicle name and login of the user who owns that chronicle. Therefore @n results in an exception if there is more than one chronicle with that name and all of them belong to different users than the current user who runs the proof script. This exception can be avoided by also specifying the login u of the user that owns the chronicle via @u:n.

It is also possible to override the current assignment of chronicles to chronicle versions and directly specify the version v via the syntax @u:n:v.

Finally, a context can be accessed directly via the syntax @"key" where key is the key of the context.

Accessing a context in one of the above ways has the potential of introducing a dependency on a context that is no ancestor of any of the contexts generated by this proof script, thus subverting intelligent versioning. To avoid this, the above ways of retrieving a context do not directly return the specified context, but rather an empty context that has been generated by this proof script and that has the specified context as its parent. This way, all dependencies are recorded via the context parent relation and thus susceptible to intelligent versioning.

**Merging of Contexts**

Contexts have been designed to model mathematical development as a linear sequence. Consequently, no method has been presented so far to *merge* two or more contexts, although the ability to move terms and theorems between different contexts already somewhat goes into that direction.

Nevertheless, merging of contexts is a concept that has been proven to be very valuable and useful in interactive theorem proving. For example, in Isabelle [4a], the notion of *theory* can be understood as a special case of the notion of context in

ProofPeer. It is possible to derive a new theory by inheriting from several other theories. This would correspond to a merge of several contexts in ProofPeer. It is easy to define a simple notion of merging context A with context B. The situation before the merge is as follows:

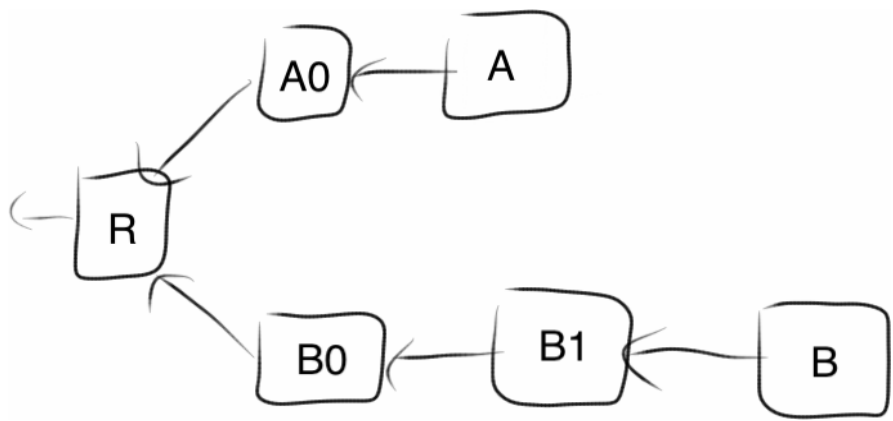

Here R is the context that has maximum depth and is ancestor of both A and B. Merging A with B could now just mean to copy and adjust the path from R to B such that it does not start at R anymore, but at A, resulting in a path A, B0', ..., B'. The result of the merge would then be B' :

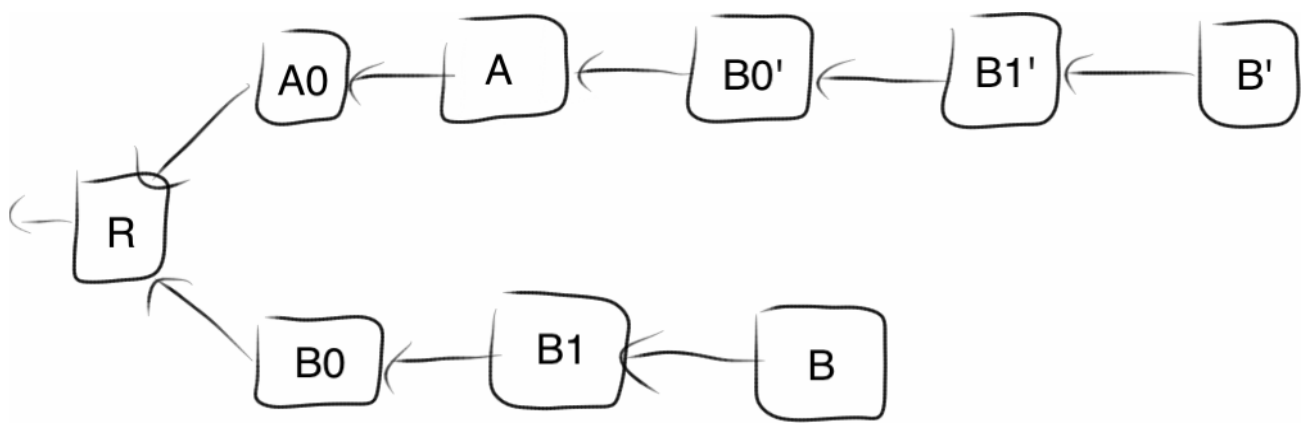

Let us call this way of merging *simple merge*. While simple merge might prove useful, it differs from the merge of Isabelle theories. Let us write the merge of theory / context A with theory / context B as $A + B$. Then let $U = A + B$ and $V = B + C$. What does $U + V$ look like? In Isabelle, the result would be $U + V = A + B + C$. In ProofPeer, simple merge yields $U + V = A + B + B + C$, resulting in duplicated contexts stemming from two copies of B instead of just one copy. Clearly, most of the time the Isabelle way of merging is preferable to simple merge.

One way of dealing with this complication could be to extend contexts with an additional field "origin" that contains the key of the original context in case the context has been the result of a merge. So in the above example, B' would contain a pointer to its origin B. When merging B' with B, the system could check that actually nothing needs to be done and that therefore B' + B = B'.

Another way of dealing with the merge problem could be to solve it not on the level of contexts, but on the level of chronicles. Chronicles are an organizational tool in ProofPeer that has similarities with the concept of theories in Isabelle, therefore it is worthwhile to investigate such a solution.

Finally, an important point that we have not looked at so far is: how should the value component V of a merged context look like? There does not seem to be a straightforward answer to this that covers all cases; it might be best to gain experience with how ProofScript is used *without* a merge facility before trying to tackle that question.

**Interactivity**

So far we have concentrated on the "theorem proving system" and the "cloud-based" part of "cloud-based interactive theorem proving system". It is time to consider the "interactive" part as well.

The most successful user interface for interactive proof is Proof General [42]. Extensions of it have been advocated and implemented [43], but have so far not found wide-spread use.

Proof General (PG) manages a proof script. The user can step through the script and undo previous steps. PG communicates with the theorem proving engine (TPE) to achieve this. The proof script is sent from PG to the TPE which breaks up the script into non-overlapping text spans. The proof state itself is solely managed by the TPE. When doing a proof step, PG informs the TPE which text span is being executed. The text spans before are marked as processed (blue color), the text spans after it are unprocessed. The TPE will return with the result that either the text span has been executed successfully, or that the execution failed. Depending on this the current text span is marked as processed, or not.

It is very desirable to have a similar interface for managing proofs in ProofPeer. The current design of ProofPeer only allows for batch processing of proof scripts: edit the proof script on the client and then ship it as one whole piece to the ProofPeer server which will execute it and display the result and output of that execution.

One problem with implementing a PG-like solution for ProofPeer is that ProofPeer's proof scripts do not entirely fit into the linear model that PG has of a proof script: a sequence of non-overlapping text spans. ProofPeer proof scripts can have if-statements, loops and recursive functions which contain proof steps. Therefore there is no linear a-priori partition of the proof script into non-overlapping lines.

Because ProofPeer's proof scripts are full programs, building an interactive user interface for ProofPeer implies building a debugger interface, something similar to the *omniscient debugger* [44]. The task is more difficult than just building a debugger for a language, though, because one should be able to edit the proof script at run time. An example of this difficulty can be observed by revisiting a previous example script:

> **have** impl = "∀ x : prop. x ⟶ x" **by**
>   **begin**
>     **fix** "x : prop"
>     **assume** h = "x"
>     **have** "x" **by** h
>   **end**

In a truly interactive proof, the script will not present itself like this, but rather like this, because the user is in the middle of the proof:

> **have** impl = "∀ x : prop. x ⟶ x" **by**
>   **begin**
>     **fix** "x : prop"

This means that the proof script parser must be able to handle such incomplete scripts, and that the prover engine must have an appropriate representation of them.

Another issue that complicates interactivity is that the evaluation of a ProofPeer proof script is done through a functional evaluator, i.e. basically a bunch of mutually recursive functions calling each other. This means that in the current form it is not possible to isolate the state of an evaluation and implement actions like step forward and undo on top of it. A solution to this problem could be to come up with an abstract machine for ProofScript evaluation so that the evaluation steps and evaluation state are explicit. Fortunately, there exists a clear road map for how to go from a functional evaluator to an abstract machine [45].

**Social Aspects of ProofPeer**

On Thomas Hales's web page [37] you will find pointed out a statement ascribed to David Dill:

> "don't rely on social processes for verification".

With this citation Hales probably wants to express that while social processes have been used for verification in Mathematics for a long time now, the availability of ITP technology makes it possible to convince yourself of a theorem without having to rely on social processes.
Of course this is to be taken with a grain of salt, as ITP technology itself may have been verified only by social processes like open source.
The ultimate goal of ProofPeer is not to replace social processes in Mathematics and Engineering, but to enrich them and make them more powerful.

The design of the first ProofPeer prototype does not contain many social features, as the focus so far has been on bringing ITP technology to the cloud. You could argue though that the ability for anyone to create a ProofPeer account, create chronicles, view anyone else's chronicles and proof scripts, and to make use of the chronicles of your peers for the creation of your own chronicles, already is a major social feature. Nevertheless, much more can and needs to be done, like the categorization of chronicles via tags, and the possibility to upvote and downvote the contributions of users. Surely much can be learnt from successful Internet communities like MathOverflow [38], StackOverflow [39], and GitHub [40].

Particularly interesting is the Polymath Project [41] that aimed to explore "massively collaborative mathematics". The project was initiated by Fields medal winner Timothy Gowers. The idea was to collaborate online through his web blog on a specific open math problem, the Hales-Jewett theorem. A few basic rules were outlined and anyone willing to adhere to these rules was invited to participate. On one hand, the project was clearly a huge success, because the problem was solved (in even more general form than originally planned) after a few weeks. On the other hand, there had been only 27 collaborators in total, so the project was hardly *massively* collaborative and would probably not scale unmodified to larger collaborations.

One of the few basic rules was to limit each comment on the blog to a single idea, even if the idea was not fully developed. While ProofPeer could potentially be very helpful with sharing a fully developed, i.e. a fully formalized, idea, it is not clear how and whether ProofPeer can be helpful for sharing not fully developed ideas. It seems that research into how to make ProofPeer helpful in that respect should turn out to be very beneficial.